\documentclass[a4paper,11pt]{article}
\usepackage{graphicx}

\pdfoutput=1

\usepackage{jheppub}
\usepackage{epsfig}
\usepackage{amssymb,amsmath}
\usepackage{setspace}
\usepackage{subfigure}
\usepackage{graphicx}
\usepackage{color}
\usepackage{cancel}

\title{Hybrid Anomaly and Gravity Mediation for Electroweak Supersymmetry}

\date{\today}

\author[1]{Bin Zhu}
\author[2]{Ran Ding}
\author[3,4]{Tianjun Li}

\affiliation[1]{Department of Physics, Yantai University, Yantai 264005, P. R. China}
\affiliation[2]{Center for High-Energy
Physics, Peking University, Beijing, 100871, P. R. China}
\affiliation[3]{Key Laboratory of Theoretical Physics and Kavli Institute for
Theoretical Physics China (KITPC), Institute of Theoretical Physics,
Chinese Academy of Sciences, Beijing 100190, P. R. China}
\affiliation[4]{School of Physical Electronics, University of Electronic Science
and Technology of China, Chengdu 610054, P. R. China}

\emailAdd{zhubin@mail.nankai.edu.cn}
\emailAdd{dingran@mail.nankai.edu.cn}
\emailAdd{tli@itp.ac.cn}

\abstract{
We propose a hybrid mediation and hybrid supersymmetry breaking. In particular the RG-invariant anomaly mediation is considered. Together with additional gravity mediation the slepton tachyon problem of anomaly mediation is solved automatically.  The special properties
are that all of color sparticles masses fall into several TeV region due to the large $m_0$ and $m_{32}$ which are well beyond the scope of current LHC Run II limits. Unlike the gauge mediation,
the dark matter candidate is still the lightest neutralino and the correct dark matter relic density can be
realized within the framework of mixed axion-wino dark matter. Due to the existence of multi-component of axion-wino dark matter, the direct detection  cross section is suppressed to evade the tightest LUX, PandaX bound. Furthermore the fine-tuning is under control when the single scale supersymmetry breaking mechanism is adapted.
}

\begin{document}

\maketitle


\section{Introduction}

Our desire to find Beyond the Standard Models (BSMs) of particle physics has largely been shaped by naturalness\cite{Martin:1997ns,Feng:2013pwa,Craig:2013cxa} and WIMP miracle argument\cite{Jungman:1995df,Hooper:2009zm}. Among various realizations of the BSMs, supersymmetry provides an elegant framework to cope with the two questions. First, the huge hierarchy between GUT scale and weak scale can be naturally stabilized through the cancellation between boson and fermion loops. Second, the lightest stable particle (LSP) realizes WIMP miracle in terms of thermal freeze-out mechanism. Supersymmetry must be softly broken\cite{Chung:2003fi} in order to distinguish the sfermions without re-introducing the hierarchy problem. There are three well-known mechanisms to generate soft terms, {\it i.e.}, anomaly mediation\cite{Chacko:1999am,Luty:2005sn}, gauge mediation\cite{Giudice:1998bp} and gravity mediation\cite{Nilles:1983ge}. The anomaly mediation is realized in extra dimensions, where SUSY breaking sector is separated from visible MSSM sector by a distance $r$ in the fifth dimension. If we assume only gravity can propagate in the bulk
while the MSSM states are confined in the $3$-branes, the interactions between two sectors can be written as follows

\begin{align}
\mathcal{L}_{\text{eff}}=\int d^4\theta\Phi^+ \exp(V)\Phi\frac{X^+X}{M^2}
+\int d^2\theta\frac{X^3}{M^3}(m\Phi^2+y\Phi^3)-\frac{i}{16\pi}\int d^2\theta
\tau W^{\alpha}W_{\alpha}+h.c.\;.
\label{eqn:effective1}
\end{align}

The Eq.~(\ref{eqn:effective1}) is associated with accidental R-symmetry with $R[X]=2/3$ and $R[\Phi]=0$. Once we integrate out the hidden sector dynamics through rescaling the superfields $\Sigma\Phi/M\rightarrow\Phi$, the effective Lagrangian becomes

\begin{align}
\mathcal{L}_{\text{eff}}=\int d^4\theta\Phi^+ \exp(V)\Phi
+\int d^2\theta\left(\frac{X}{M}m\Phi^2+y\Phi^3\right)-\frac{i}{16\pi}\int d^2\theta
\tau W^{\alpha}W_{\alpha}+h.c.\;.
\label{eqn:effective2}
\end{align}

It is easy to find that when $m$ is set to be zero, the Lagrangian in Eq.~(\ref{eqn:effective2})
is classically scale invariant. Nevertheless, the scale invariance is broken by quantum corrections\cite{ArkaniHamed:2001tb}.
After considering the quantum corrections and renormalizing the effective theory down to scale $\mu$, we have
\begin{align}
\mathcal{L}_{\text{eff}}=\int d^4\theta Z\left(\frac{\mu M}{\Lambda X},\frac{\mu X}{\Lambda X^+}\right)\Phi^+ \exp(V)\Phi
+\int d^2\theta y\Phi^3-\frac{i}{16\pi}\int d^2\theta
\tau W^{\alpha}W_{\alpha}+h.c.\;.
\label{eqn:effective3}
\end{align}

The wavefunction renormalization $Z$ is real and R-symmetric, it must have the function of $(\mu M/\Lambda\left|\Sigma\right|)$. In addition, the kinetic coupling $\tau$ is holomorphic, so it must have the following form
\begin{align}
\tau=i\frac{\tilde b}{2\pi}\ln\left(\frac{\mu M}{\Lambda X}\right)\;,
\end{align}
where the dependence of $\mu$ determines that $\tilde b=b$. Supersymmetry breaking effect is
communicated to auxiliary supergravity fields which induces a gaugino masses

 \begin{align}
 M_{\lambda}=\frac{i}{2\tau}\frac{\tau}{X}F_{X}=\frac{bg^2}{16\pi^2}\frac{F_{X}}{M}\;.
 \end{align}
Because the SUSY breaking mass arises through the one-loop anomaly, this mechanism is dubbed
as anomaly mediation. We can also expand the wavefunction renormalization $Z$ in superspace
 \begin{align}
 Z=\left[Z-\frac{1}{2}\frac{\partial Z}{\partial\ln\mu}\left(\frac{F}{M}\theta^2+\frac{\bar F}{M}\bar\theta^2\right)+\frac{1}{4}\frac{\partial^2 Z}{\partial\ln\mu^2}\frac{F^2}{M^2}\theta^2\bar\theta^2\right]\;.
 \end{align}

In terms of the definition of anomalous dimension and beta function, we obtain
\begin{align}
Z\Phi\exp(V)\Phi=\left[1+\frac{1}{4}
\left(\frac{\partial\gamma}{\partial g}\beta_g+\frac{\partial\gamma}{\partial y}\beta_y\right)\frac{F^2}{M^2}\theta^2\bar\theta^2\right]\Phi^+\exp(V)\Phi\;.
\end{align}

The squark and slepton masses are easy to identify
\begin{align}
m_{\tilde\Phi}^2=-\frac{1}{4}\left(\frac{\partial\gamma}{\partial g}\beta_g
+\frac{\partial\gamma}{\partial y}\beta_y\right)\;.
\label{eqn:gaugino}
\end{align}

At one-loop order, we can get some feelings about the soft terms with,
\begin{align}
\gamma&=\frac{1}{16\pi^2}(4C_2[r]g^2-ay^2)\;,\nonumber\\
\beta_g&=-\frac{bg^3}{16\pi^2}\;,\nonumber\\
\beta_y&=\frac{y}{16\pi^2}(ey^2-fg^2)\;,
\end{align}
As a consequence,
\begin{align}
m_{\tilde\Phi}^2=\frac{1}{512\pi^4}(4C_2[r]bg^4+ay^2(ey^2-fg^2))\frac{F^2}{M^2}\;.
\label{eqn:sfermion}
\end{align}

It is easy to see that anomaly mediated SUSY breaking takes an elegant form to generate soft terms, in which the soft terms in Eq.~(\ref{eqn:gaugino}) and (\ref{eqn:sfermion}) are determined by the appropriate power of the gravitino mass multplied by perturbatively calculable functions of anomalous dimensions and beta functions. In other words, these soft terms are renormalization group (RG) invariant. Furthermore, the sfermion masses as well as trilinear soft terms are just power series in the Yukawa matrices, so it is called the minimal flavor violation scenario (MFV). Therefore, the flavor violation is suppressed greatly as the SM did. However the problem appears when we consider the sfermion mass seriously. For squark masses they are always positive due to asymptotically free gauge theories. The fatal issues of anomaly mediation
exists when we find the tachyonic slepton masses in which the gauge coupling
in Eq.~(\ref{eqn:sfermion}) is not asymptotically free.

There are lots of approaches to stabilize the tachyonic slepton masses. For example, we can introduce new bulk superfields to couple lepton and spurion $X$ which gives rise to additional contribution to slepton masses. Another approach is to consider new Higgs doublet with large Yukawa couplings. In this setup the large Yukawa couplings is used to cancel the negative $U(1)_Y$ and $SU(2)_L$ contributions. Besides the two approaches, the most well-known approach is to include heavy SUSY violating threshold effects such as gauge mediation. The combination of gauge and anomaly mediation seems to provide an elegant framework to study the MSSM phenomenology and is denoted by hybrid mediation or mirage mediation\cite{Choi:2006xb,Abe:2007je,Everett:2008qy,Everett:2008ey,Baer:2016hfa} for simplicity. The shortcoming of this framework is that it not only breaks the RG-invariance of anomaly mediation but can not account for the $\mu$ problem. This strongly suggests us to consider another possibility of hybrid mediation\cite{Zhu:2016ncq}, where gravity mediation is re-introduced to stabilize the slepton masses. Furthermore, it solves the $\mu$ problem by Giudice-Masiero mechanism\cite{Giudice:1988yz}.
The price we pay is the dangerous flavor violation constraint. We have to add the $m_0$ to the mass matrices diagonally\cite{Rattazzi:1999qg,Paige:1999ui} in order to escape the FCNC constraints. The hybrid mediation leads to a distinct and constrained particle spectrum, in which wino becomes lightest stable particle (LSP). The more complicated story comes when we consider the fine-tuning. The lower fine-tuning favors small $\mu$ term in the model. As a consequence, we can have two different cases for LSP:
\\
$(\textbf{1})$ LSP is pure Wino,
\\
$(\textbf{2})$ LSP is mixed Wino-Higgsino.
\\
The properties of DM strongly depends on its constitutes. If the DM is associated with $SU(2)$ representation, i.e., wino/higgsino, relic density constraint prefers heavy LSP due to their efficient annihilation. To be specific, thermally-produced wino (higgsino) is required to be around $2.5$ ($1$) TeV in order to provide the observed DM abundance~\cite{ArkaniHamed:2006mb,Hisano:2006nn}. On the other hand, the Bino dark matter favors light sfermions which are almost excluded by the no sign of new physics at LHC. The general mixed case such as Wino-Higgsino DM are strongly constrained from the direct detection. Therefore, for the case $(\textbf{1})$, it is impossible to obtain correct relic density with light wino. While the case $(\textbf{2})$ is ruled out by the direct detection. In order to construct a model with light wino DM, we then appeal to mixed axion-neutralino DM scenario to fill this gap. This issue will be discussed in details in Sec.~\ref{sec:mixedDM}. Due to the introduction of mixed axion-wino Dark Matter scenario the direct detection cross section is suppressed by a factor $\Omega h_{\tilde w}^2/0.11$ which easily evade the current LUX constraint.

This paper is organized as follows: In Sec.\ref{sec:soft} we give a overview of soft terms. In Sec.\ref{sec:Phe} the phenomenology of the model is discussed in ddetail in particular the relic density as well as its direct detection cross section are explored. It shows that the parameter space survives even when we consider the wino LSP scenario.

\section{Soft Terms}
\label{sec:soft}
The soft terms in anomaly mediation is easy to identify in Eqs.~(\ref{eqn:gaugino}) and
 (\ref{eqn:sfermion}). Explicitly, the gaugino masses are given by
\begin{align}
M_1&=\frac{33}{5}\frac{g_1^2m_{32}}{16 \pi ^2}\sim\frac{m_{32}}{120}\;,\nonumber\\ M_2&=\frac{g_2^2 m_{32}}{16 \pi ^2}\sim\frac{m_{32}}{360}\;,\nonumber\\
M_3&=-3\frac{g_3^2 m_{32}}{16 \pi ^2}\sim-\frac{m_{32}}{40}\;.
\label{eqn:gaugino2}
\end{align}
with the sfermion masses
\begin{align}
m_{u_3}^2&=\frac{m_{32}^2 \left(-\frac{88 g_1^4}{25}+8 g_3^4+2 \beta
   _t y_t\right)}{256 \pi ^4}\;,\nonumber\\
m_{d_3}^2&=\frac{m_{32}^2 \left(2 \beta _b y_b-\frac{22 g_1^4}{25}+8
   g_3^4\right)}{256 \pi ^4}\;,\nonumber\\
m_{q_3}^2&=\frac{m_{32}^2 \left(\beta _b y_b-\frac{11
   g_1^4}{25}-\frac{3 g_2^4}{2}+8 g_3^4+\beta _t
   y_t\right)}{256 \pi ^4}\;,\nonumber\\
m_{L_3}^2&=\frac{m_{32}^2 \left(-\frac{99 g_1^4}{50}-\frac{3
   g_2^4}{2}+\beta _{\tau } y_{\tau }\right)}{256 \pi ^4}\;,\nonumber\\
m_{e_3}^2&=\frac{m_{32}^2 \left(2 \beta _{\tau } y_{\tau }-\frac{198
   g_1^4}{25}\right)}{256 \pi ^4}\;,\nonumber\\
m_{H_d}^2&=\frac{m_{32}^2 \left(-\frac{99 g_1^4}{50}-\frac{3
   g_2^4}{2}+\beta _{\tau } y_{\tau }+3\beta _{b} y_{b}\right)}{256 \pi ^4}\;,\nonumber\\
m_{H_u}^2&=\frac{m_{32}^2 \left(-\frac{99 g_1^4}{50}-\frac{3
   g_2^4}{2}+3\beta _{t} y_{t}\right)}{256 \pi ^4}\;.
\label{eqn:sfermion2}
\end{align}
In spite of gauge mediation, there will be additional trilinear soft terms induced by anomaly mediation,
\begin{align}
T_{ijk}=\frac{1}{2}(\gamma_i+\gamma_j+\gamma_k)y_{ijk}\frac{F}{M}~.~
\label{eqn:trilinear}
\end{align}

From Eqs.~(\ref{eqn:gaugino2}),~(\ref{eqn:sfermion2}) and (\ref{eqn:trilinear}), it is easy to find that anomaly is insensitive to the UV physics. In other words the RG-invariance enables us to add anomaly mediation at any energy scale. As a consequence the phenomenology of the model is completely determined by the low energy effective theory. Such a nice property of UV insensitivity cannot be  retained when we consider the tachyonic slepton masses. In order to remove the unpleasant fact, we introduce conventional gravity mediation $m_0$ to lift the sleptons. Certainly we should also include $m_{12}$ and $A_0$ to overwrite the anomaly mediation:

\begin{itemize}
\item The distinguished prediction of anomaly mediation is wino LSP. When we introduce $m_{12}$ to gaugino masses, the wino LSP scenario will be destroyed. In order to maintain the anomly mediation to be dominated, $m_{12}$ must be ignored.

\item Though anomaly mediation contains non-vanishing trilinear soft terms, $125$ GeV higgs requires even larger $A_t$ in order to obtain reliable fine-tuning. That is why we keep $A_0$ being input parameter.
\end{itemize}

 Finally we find gravity mediation not only provides $m_0$ but solves the $\mu$ problem naturally. We only leave following input parameters,
\begin{align}
\{m_0,m_{32},\tan\beta,\text{Sign}(\mu), A_0\}
\end{align}

\section{Phenomenology}
\label{sec:Phe}

In this section, we present the numerical results for hybrid mediation models which include the fine-tuning measure and dark matter properties. In our numerical analysis, the relevant soft terms are firstly generated at GUT scale in terms of gravity mediation and anomaly mediation. The low scale soft terms are obtained by soving the two-loop RG equations. For this purpose, we implemented the corresponding boundary conditions in Eqs.~(\ref{eqn:gaugino2}), (\ref{eqn:sfermion2}) and (\ref{eqn:sfermion2}) into the Mathematica package {\tt SARAH}~\cite{Staub:2008uz,Staub:2009bi,Staub:2010jh,Staub:2012pb,Staub:2013tta}. Then {\tt SARAH} is used to create a {\tt SPheno}~\cite{Porod:2003um,Porod:2011nf} version for the MSSM to calculate particle spectrum, and {\tt micrOMEGAs}~\cite{Belanger:2013oya} for the evaluation of the
density and direct detction cross sections of dark matter. The tasks of parameter scans are implemented by package~{\tt SSP}~\cite{Staub:2011dp}.

\begin{figure}[!htbp]
\begin{center}
\includegraphics[width=0.7\linewidth]{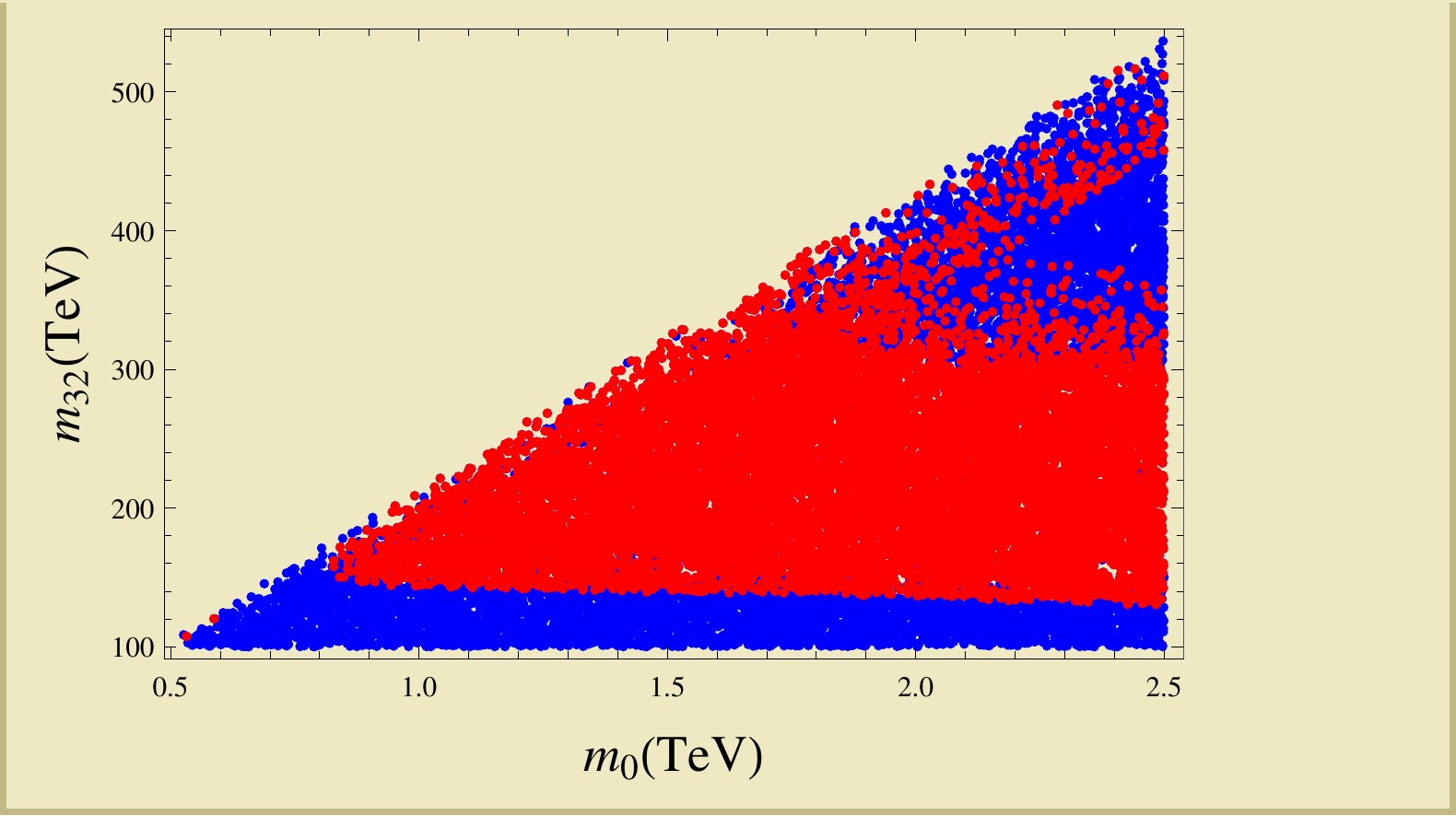}
\end{center}
\caption{The distributions of samples in $[m_{32},m_0]$ plane. Here the blue (red) points denote total (constrained) samples.}
\label{fig:tachyon}
\end{figure}

\begin{figure}[!htbp]
\begin{center}
\includegraphics[width=0.7\linewidth]{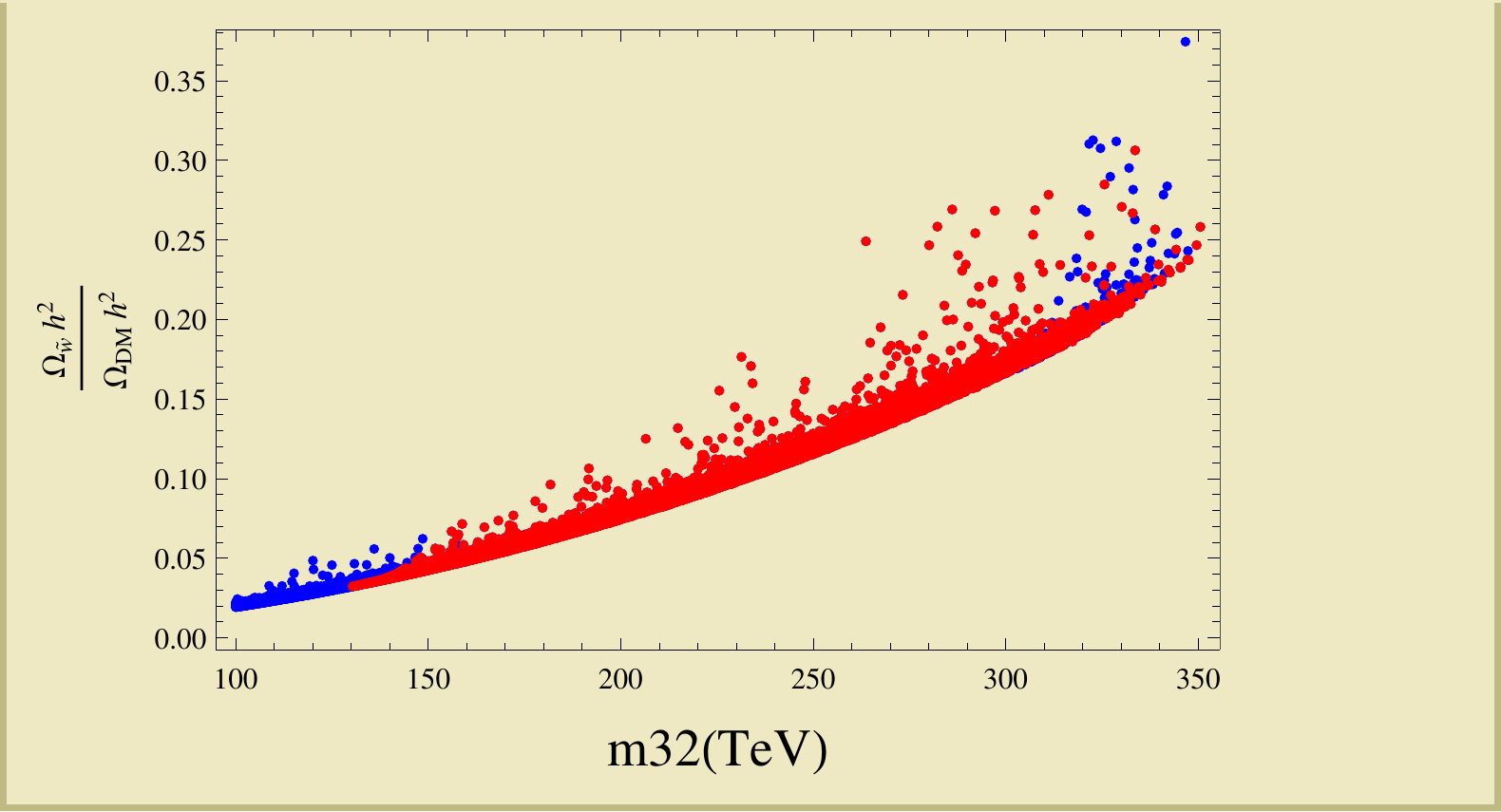}
\end{center}
\caption{Wino thermal abundance fraction $\Omega_{\tilde W}h^2/\Omega_{\rm DM}h^2$ versus $m_{32}$.}
\label{fig:relic}
\end{figure}

\begin{figure}[!htbp]
\begin{center}
\includegraphics[width=0.7\linewidth]{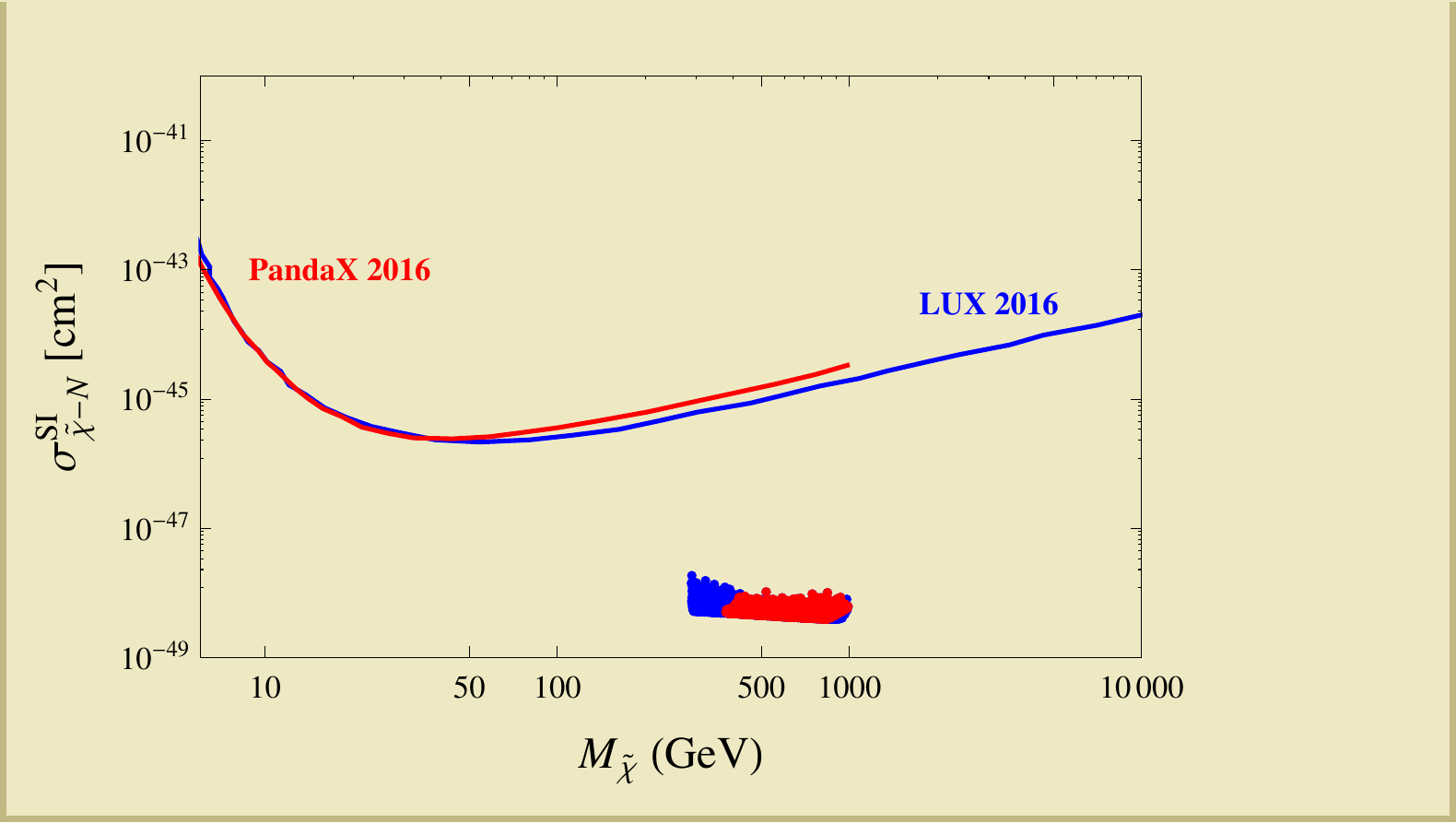}
\end{center}
\caption{Spin independent wino-nucleon cross section as a function of wino DM mass. For comparison, the latest exclusion limits from LUX and PandaX-II experiments are also shown.}
\label{fig:direct}
\end{figure}

We implement a random scan in our parameter space within following ranges:
\begin{align}
m_0\in[500,~2500]~{\rm GeV}\;,\quad m_{32}\in[10^5,~10^6]~{\rm GeV}\;, \tan\beta\in[10,30],\; A_0\in[-6000,6000]
\label{eq:ranges}
\end{align}
and fix ${\rm sign}(\mu)=1$. During the scan, various mass spectrum and low energy constraints have been considered and listed at below:
\begin{enumerate}
  \item {The higgs mass constraints:
\begin{align}
 123 {\rm GeV}\leq m_h \leq 127 {\rm GeV}\;,
\end{align}}
  \item {LEP bounds and B physics constraints:
\begin{eqnarray}\label{eqn:Bphysics}
1.6\times 10^{-9} \leq{\rm BR}(B_s \rightarrow \mu^+ \mu^-)
  \leq 4.2 \times10^{-9}\;(2\sigma)~\text{\cite{CMS:2014xfa}}\;,
\nonumber\\
2.99 \times 10^{-4} \leq
  {\rm BR}(b \rightarrow s \gamma)
  \leq 3.87 \times 10^{-4}\;(2\sigma)~\text{\cite{Amhis:2014hma}}\;,
\nonumber\\
7.0\times 10^{-5} \leq {\rm BR}(B_u\rightarrow\tau \nu_{\tau})
        \leq 1.5 \times 10^{-4}\;(2\sigma)~\text{\cite{Amhis:2014hma}}\;.
\end{eqnarray}}
  \item {Sparticle bounds from LHC Run-II:
\begin{itemize}
  \item Light stop mass $m_{{\tilde t}_1} > 850$ GeV~\cite{ATLAS:2016jaa,CMS:2016inz},
  \item Light sbottom mass $m_{{\tilde b}_1}>840-1000$  GeV~\cite{Aaboud:2016nwl,CMS:2016xva},
  \item Degenerated first two generation squarks (both left-handed and right-handed) $m_{{\tilde q}}>1000-1400$ GeV~\cite{CMS:2016xva},
\item Gluino mass $m_{\tilde g} > 1800$ GeV~\cite{ATLAS:2016kts,CMS:2016inz}.
\end{itemize}}
\end{enumerate}
The samples which are satisfied all above constraints are denoted as constrained samples. We display the most representative parameters space and in Figs.~\ref{fig:tachyon}-\ref{fig:direct}. For all of figures, blue (red) points denote total (constrained) samples. The distributions of samples in $[m_{32},m_0]$ plane are shown in Fig.~\ref{fig:tachyon}. One can see that blank area in the left-top corresponds to the invalid parameter space which is resulted from tachyonic sleptons. Moreover, the constraints from LHC direct SUSY searches and low-energy observations are generally easy to be satisfied, thus, the valid parameter space primarily determined by higgs mass requirement. To obtain correct higgs mass, it is then impose a lower bounds with $(m_{32},~m_0)>(140,~0.8)$ TeV. In the following two subsections, we will explore the DM and fine-tuning properties in our
model. Finally we show the benchmark points of our model

\begin{table*}[hbtp]
\begin{tabular}{|c|c|c|c|c|c|c|c|c|}
\hline
Parameters & $m_{32}$ & $m_0$ & $A_0$ & $\tan\beta$ & $m_h$ & $m_{{\tilde t}_1}$ & $m_{\tilde g}$ & $\Delta _{FT}$   \\\hline
Output & $1.7\times10^5$ & $1995$ & $-6650$ & $10$ & $124$ & $2678$ & $3496$ & $2057$ \\\hline
\end{tabular}
\caption{The input parameters, important particle spectra and $\Delta_{FT}$ for benchmark points. }
\label{tab:ranges}
\end{table*}

\subsection{Mixed axion-wino dark matter}
\label{sec:mixedDM}

From equation (\ref{eqn:gaugino2}), one expects the ratio of gaugino masses at weak-scale yield $M_1:M_2:M_3\sim3:1:8$, which then indicating a wino LSP as the thermal DM candidate. It is well known that the typical thermally produced relic density of wino LSP yields~\cite{ArkaniHamed:2006mb,Hisano:2006nn}
\begin{align}
\Omega_{\tilde W}h^2\sim0.12(M_2/2.5{\rm TeV})^2\;.
\end{align}
In order to obtain the correct relic density while keep the light WIMP DM at the same time, one can introduce non-WIMP component to saturate relic abundance. One of interesting solution is the mixed axion-wino DM scenario~\cite{Bae:2014rfa,Baer:2014eja,Bae:2015rra}. The original motivation of the axion is to solve the strong CP problem in the QCD sector of SM. Since the QCD $\theta$ vacuum does not respect $U(1)_A$ symmetry~\cite{Weinberg:1975ui}, the QCD lagrangian contains a CP-violating term $\bar{\theta}\frac{g^2_3}{32\pi}G_{a\mu\nu}{\tilde G}^{a\mu\nu}$~\cite{tHooft:1976rip}, the requirement of extremely small $\bar{\theta}$ then inducing large fine-tuning on $\bar{\theta}$. This is so called strong CP problem and is solved by the Peccei-Quinn (PQ) mechanism~\cite{Peccei:1977hh}. Such CP-violating term then dynamically tends to zero when $U(1)_{\rm PQ}$ is broken, and the corresponding Nambu-Goldstone boson is the axion~\cite{Weinberg:1977ma,Wilczek:1977pj}. In the axion extend MSSM, the axion superfield is defined as
\begin{align}
A=\frac{1}{\sqrt{2}}(s+ia)+\sqrt{2}\theta{\tilde a}+\theta^2F\;,
\end{align}
where $a$, $\tilde a$ and $s$ are respectively denote the axion, axino and saxion fields. In gravity mediation, $m_a$ and $m_s$ are both expected
to be of order of $m_{32}$. In the absence of couple to matter, axion and saxion can be produced via coherent oscillations due to misalignment mechanism ~\cite{Preskill:1982cy,Abbott:1982af,Dine:1982ah,Turner:1985si,Lyth:1991ub,
Visinelli:2009zm}, which is totally determined by axion mass $m_a$ and axion decay constant $f_a$. While in the mixed axion-neutralino DM case, DM is composed of both WIMPs (wino) and axions. In such case, one should take into account the following effects during the calculation of relic abundance~\cite{Bae:2014rfa}:
\begin{itemize}
  \item {In addition to usual thermal production, WIMPs can also be produced through production and subsequent decay of both axinos and saxions in the early Universe.}
  \item {Any existing relics can be diluted by inject late-time entropy into the early Universe resulted from saxions production via coherent oscillations. }
  \item {Finally, except that usual coherent production, axions can also be thermally produced through axion-WIMP interactions and through saxion decay.}
\end{itemize}
Based on above reasons, the calculation of mixed axion-neutralino DM relic density involves numerical solution of series of coupled Boltzmann equations. This issue has been investigated carefully in Ref.~\cite{Bae:2014rfa,Bae:2015rra} for two well known axion models, i.e., SUSY KSVZ~\cite{Kim:1979if,Shifman:1979if} and SUSY DFSZ~\cite{Dine:1981rt,Zhitnitsky:1980tq} model. The detailed calculation has beyond the scope of this paper. Here we emphasize that according to their conclusion, for any remaining wino DM abundance, the resulting axion abundance in general can be adjusted to compensate the budget. This advantage make us can always saturate the observed relic abundance. We display Wino thermal abundance fraction $\Omega_{\tilde W}h^2/\Omega_{\rm DM}h^2$ versus $m_{32}$ in figure~\ref{fig:relic}. In the figure, $\Omega_{\rm DM}h^2$ denotes observed relic abundance, here we adopt the central value of combined measurement from Planck Collaboration ($68\%$ limits, Planck+WP+highL+BAO): $\Omega_{\rm DM}h^2=0.1187$~\cite{Ade:2013zuv}. As one expects, the wino relic fraction falls into the ranges from $0.02$ to $0.32$ for our interested parameter space, and monotonously increases with $m_{32}$ since $\Omega_{\tilde W}h^2\sim M^2_2\sim m^2_{32}$. Moreover, the constrained samples can be realized in wide parameter ranges. In figure~\ref{fig:direct}, we plot spin independent wino-nucleon cross section as a function of wino DM mass. For comparison, the latest exclusion limits from LUX~\cite{Akerib:2016vxi} and PandaX-II~\cite{Tan:2016zwf} Collaborations are also shown. Notice that the fraction for each DM component in the
local DM density is same as its fraction in relic abundance and the axion component has no effect on direct detection, the scattering cross section should also be rescaled by the factor $\Omega_{\tilde W}h^2/\Omega_{\rm DM}h^2$. One can see that the cross section is much lower than current direct detection limits. As a consequence, in the framework of mixed axion-wino DM scenario, our model can safely evade stringent constraints from direct detection while accord with measured relic
abundance.

\subsection{Fine-tuning and super-natural supersymmetry}

\begin{figure}[!htbp]
\begin{center}
\includegraphics[width=0.7\linewidth]{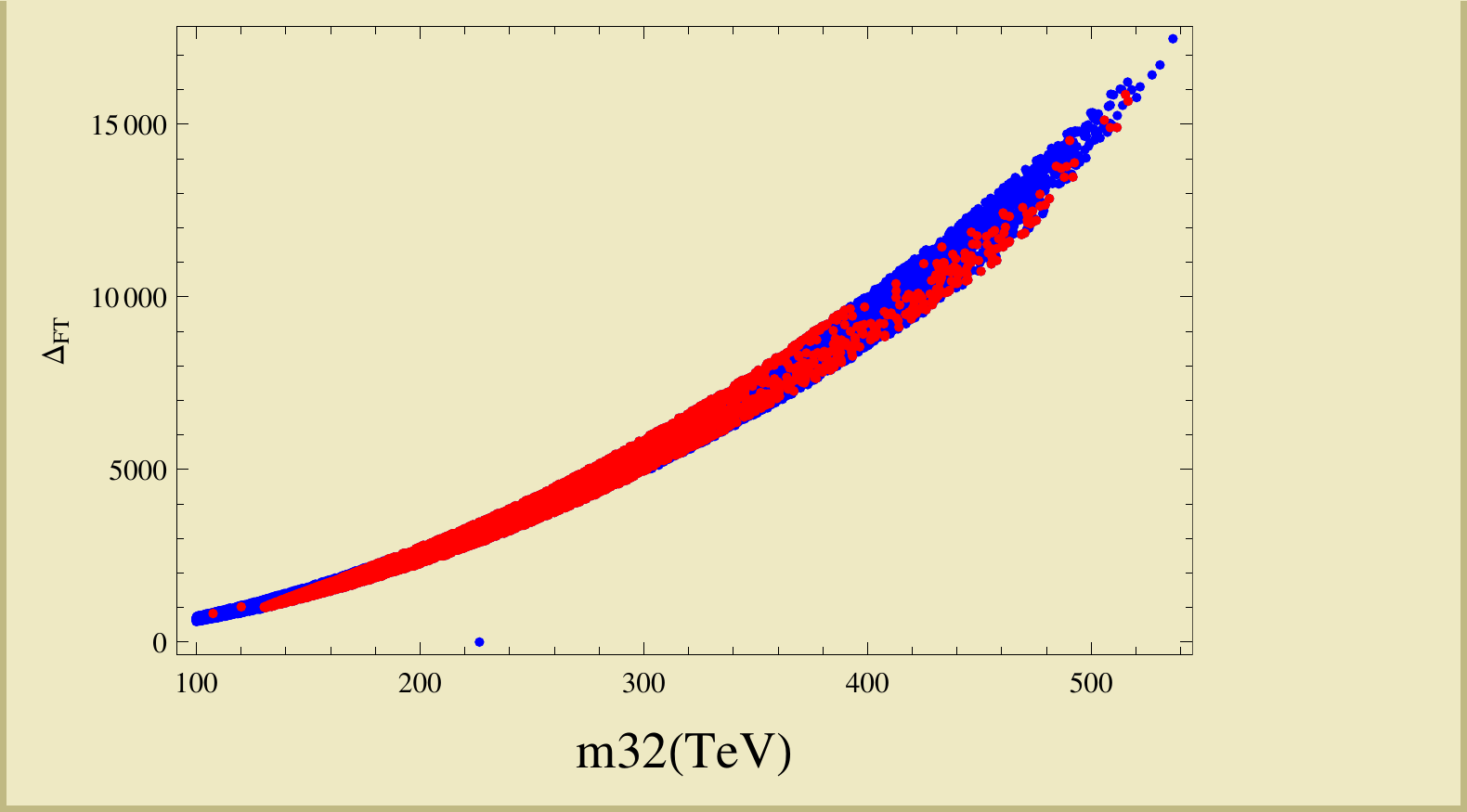}
\end{center}
\caption{Fine-tuning measure $\Delta_{\text{FT}}$ versus $m_{32}$.}
\label{fig:finetuning}
\end{figure}

\begin{figure}[!htbp]
\begin{center}
\includegraphics[width=0.6\linewidth]{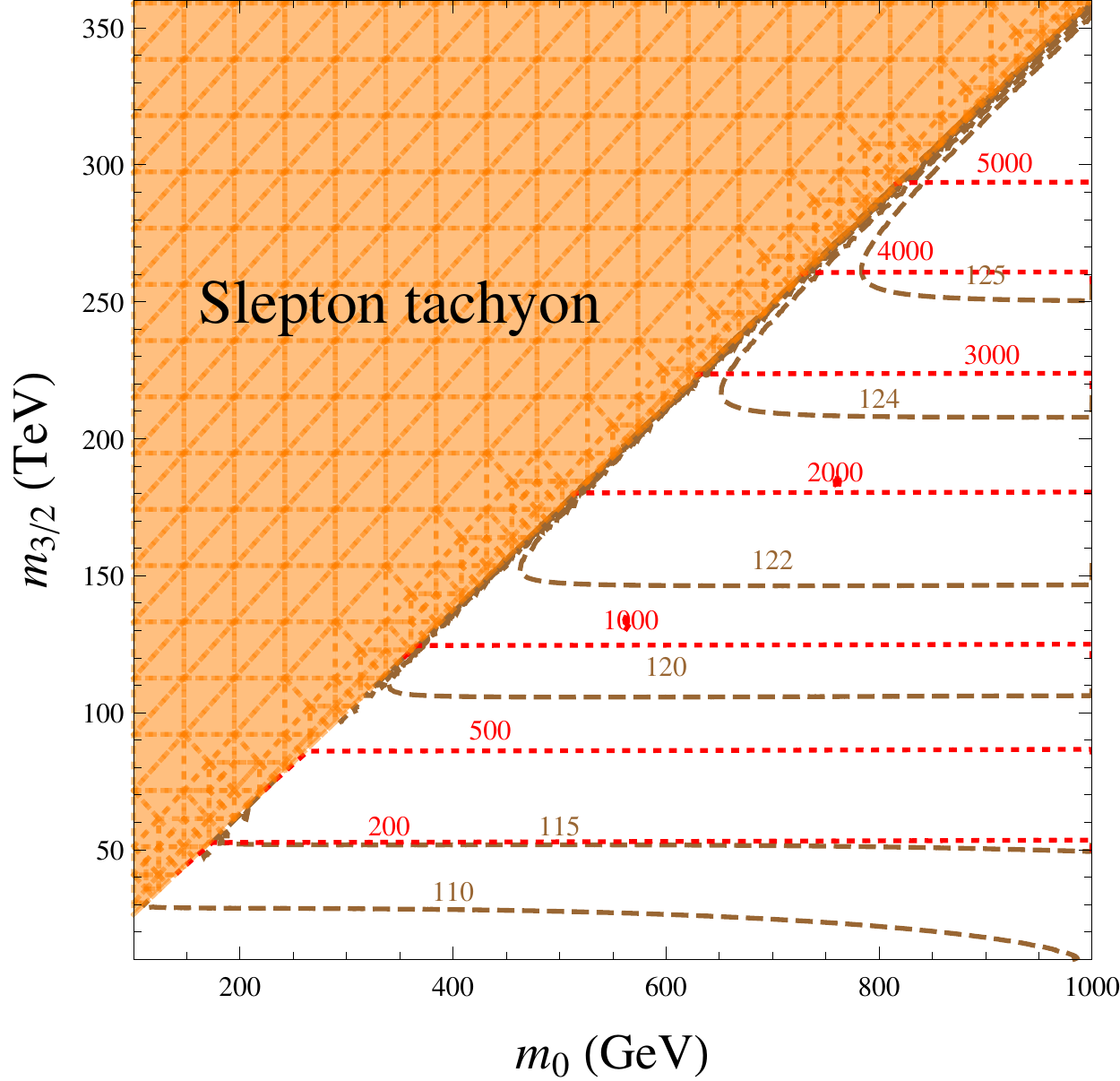}
\end{center}
\caption{The distributions of samples in $[m_{32},m_0]$ plane. The red dashed contour stands for the conventional fine-tuning measure. The brown dashed contour corresponds to the higgs mass. For a moderate higgs mass $m_h\sim 124\text{GeV}$, we find the fine-tuning is around $3000$. }
\label{fig:higgsF}
\end{figure}

\begin{figure}[!htbp]
\begin{center}
\includegraphics[width=0.6\linewidth]{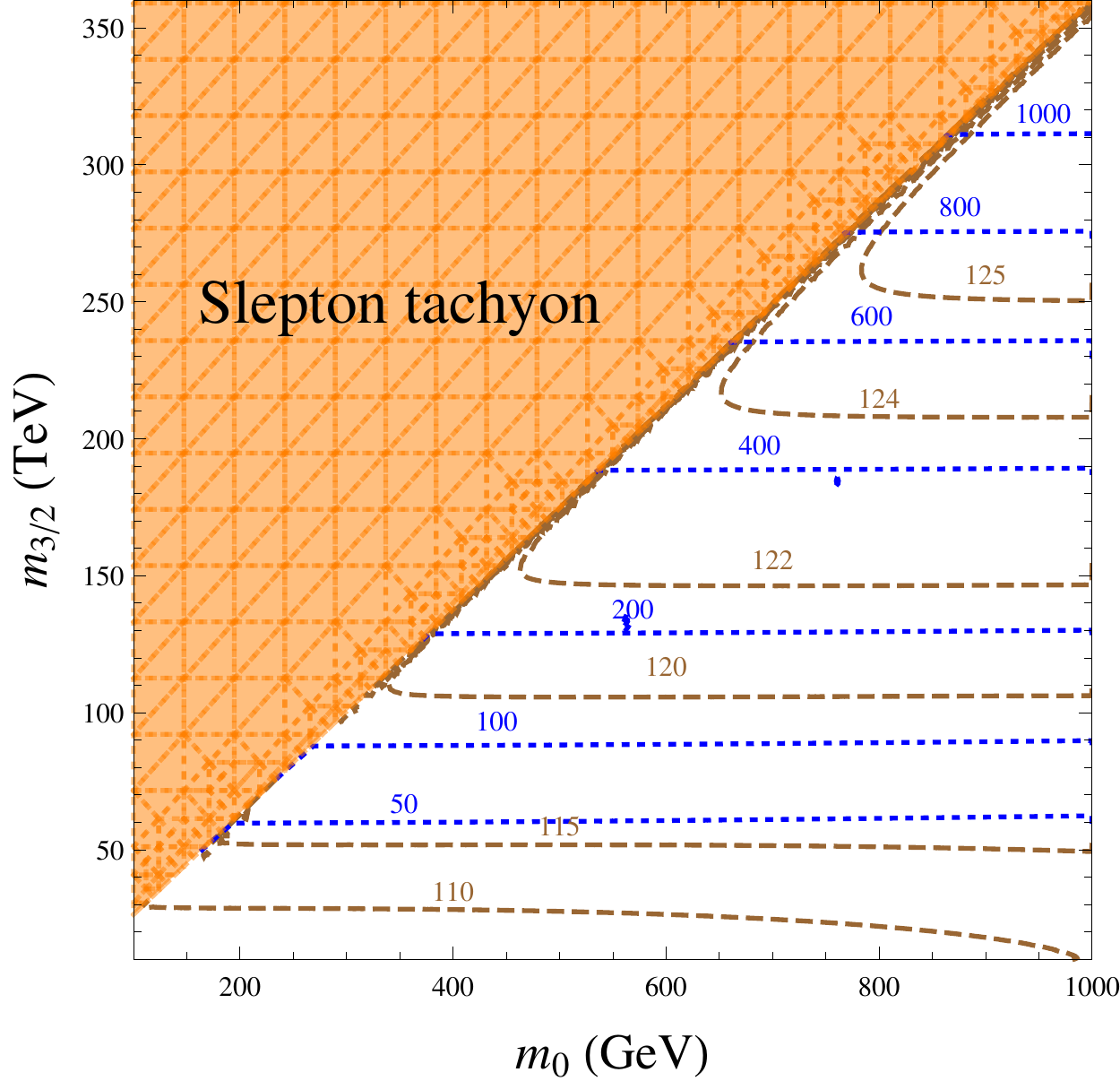}
\end{center}
\caption{The distributions of samples in $[m_{32},m_0]$ plane. The blue dashed contour stands for the super-natural fine-tuning measure induced by single scale supersymmetry breaking. The brown dashed contour corresponds to the higgs mass. For a moderate higgs mass $m_h\sim 124\text{GeV}$, we find the fine-tuning is around $600$. }
\label{fig:higgsS}
\end{figure}

There are in general four dimensional parameters in our construction: $m_0$, $m_{32}$, $\mu$ and $B_{\mu}$ which are the origin of the fine-tuning. Through the definition of Barbieri-Giudice fine-tuning measure~\cite{Barbieri:1987fn}, we can quantitively calculate the derived fine-tuning of the model,
\begin{align}
\Delta_{\text{FT}}=\text{Max}\left\{\Delta_{\alpha}\right\},\;
\Delta_{\alpha}=\frac{\partial\ln M_Z^2}{\partial\ln \alpha^2}~,~\,
\end{align}
where $\alpha$ denotes for the independent parameters as we are concerned, and $\Delta_{\alpha}^{-1}$ gives an rough estimate of the accuracy to which the parameter $\alpha$ must be tuned in order to get the correct electroweak symmetry breaking. For large regions in the parameter space with desire higgs mass, the main EWPT sources come from $\mu$ and $m_{32}$. That is mainly because large fine-tuning comes from unnatural cancellation between $\mu$ and $m_{H_u}^2$ when we solve the tadpole equations. In addition $m_{H_u}^2$ are mainly determined by the boundary condition and stop running. As a consequence large $m_{32}$ will induce large $m_{H_u}^2$. In figure~\ref{fig:finetuning} we show the dependence between $m_{32}$ and $\Delta_{FT}$. It is easy to find that the overall fine-tuning increases quickly with increasing values for $m_{32}$. The genearl value of fine-tuning is around $3000$. Even in the best point with large $A_0$, the fine tuning can be reduced to $2000$ in table~\ref{tab:ranges}. The same behavior happens when we consider the $\mu$ with the constraint of higgs mass. For naturalness we can take the following attitudes:

\begin{itemize}
\item Like dark energy we have no good ideas on how to explain the $2000$ fine-tuning based on symmetry principle or dynamical process. Therefore landscape framework is proposed to solve the problem.
\item Compared with quardratic divergence, the little fine-tuning around $2000$ is acceptable. In other words in terms of some delicated model construction, the little hierarchy can be improved. For example, through introducing interactions between messenger and higgs the fine-tuning can be reduced to $2000$\cite{Draper:2011aa,Knapen:2013zla,Evans:2013kxa,Craig:2013wga,Craig:2012xp,Ding:2013pya,Ding:2014bqa,Ding:2015vla}.
\item Resort to special approach for almost vanishing fine-tuning. Since the fine-tuning is quantified by Barbieri-Giudice fine-tuning measure~\cite{Barbieri:1987fn}, it can be vanishing for some mathamatical reasons. The well-known approach includes focus point supersymmetry\cite{Feng:1999mn,Feng:1999zg,Feng:2000bp,Feng:2011aa,Feng:2012jfa} and single-scale supersymmetry\cite{Ding:2015epa}.
\end{itemize}
In this paper we take the second approach. In terms of large A-term, the fine-tuning is reduced to the accepable level. The point is that if we want to obtain even smaller fine-tuning, the third approach is inevitable. The study of focus point supersymmetry in hybrid mediation is beyond of the paper. We leave it in the future work. For now we content ourselves within single-scale supersymmetry i.e. supernatural where the situation changes with the assumption that all the dimensional parameters are correlated at the GUT scale,
\begin{align}
m_{0}\sim m_{32}\sim \mu\sim\sqrt{B_{\mu}}
\label{eqn:natural}
\end{align}
The assumption in Eq.~\ref{eqn:natural} is reasonable when the anomaly mediation and gravity mediation come from the same supergravity theories. Furthermore the $\mu$ and $B_{\mu}$ parameters are generated by Giudice-Masiero mechanism which can be thought of the same source. Within the physical assumption that there is only one fundamental parameter with dimension mass at hand, the fine-tuning can be greatly improved with nearly vanishing fine-tuning. The conjecture is proven in the numerical calculation as we have done in \cite{Ding:2015epa}. As we know for a single scale supersymmetry breaking the conventional fine-tuning measure is no long valid, we propose a super-natural fine-tuning measure which can be implemented into SPheno easily.
The figure~\ref{fig:higgsF} just reproduces the known fact that a relatively heavy higgs $(124 \text{GeV})$ leads to percent level fine-tuning. However it is demonstrated in figure~\ref{fig:higgsS} that the super-natural fine-tuning measure quantifies very tiny fine-tuning compared with conventional one. Therefore we can not only obtain a reliable dark matter scenario in hybrid mediation but obtain a very tiny fine-tuning.

\section{Conclusion}
\label{sec:conclusion}
The wino dark matter in anomaly mediation is threatened by the severe direct detection. In terms of mixed axion-wino dark matter scenario, the relic density is easily realized. Furthermore since the axion plays no role in nuclean-dark matter scattering, the direct detection cross section is  rescaled by the factor $\Omega_{\tilde W}h^2/\Omega_{\rm DM}h^2$. As a consequence, in the framework of mixed axion-wino DM scenario, our model can safely evade stringent constraints from direct detection while accord with measured relic
abundance.

The fine-tuning of model is around $2000$ which can be accepted like extended gauge mediation. The point is that we can obtain very tiny fine-tuning in terms of single-scale supersymmetry breaking. Finally the higgs mass is easily realized and sparticles satisfy the LHC bounds.

\bibliographystyle{ArXiv}
\bibliography{Anomaly}

\end{document}